\documentstyle[multicol,aps,prb,epsf]{revtex}
\newcommand{\Vec}[1]{\mbox{\boldmath$#1$}}
\begin{document}

\draft

\title{Phase diagram of the two-dimensional extended Hubbard model:
Phase transitions between different pairing symmetries when charge and
spin fluctuations coexist}

\author{
Seiichiro Onari, Ryotaro Arita, Kazuhiko Kuroki$^1$, and Hideo Aoki
}

\address{Department of Physics, University of Tokyo, Hongo,
Tokyo 113-0033, Japan}
\address{$^1$Department of Applied Physics and Chemistry,
University of Electro-Communications, Chofu, Tokyo 182-8585, Japan}

\date{\today}

\maketitle

\begin{abstract}
In order to explore how superconductivity arises when 
charge fluctuations and spin fluctuations coexist, 
we have obtained a phase diagram against the off-site repulsion 
$V$ and band filling $n$ for the extended, repulsive 
Hubbard model on the square lattice with the fluctuation exchange 
approximation. 
We have found the existence of (i) a transition between 
$d_{xy}$ and $d_{x^2-y^2}$ pairing symmetries, 
(ii) $f$-pairing in between the $d_{x^2-y^2}$ and 
CDW phases for intermediate $0.5<n<1.0$ and large $V$, and 
(iii) for anisotropic cases the pairing symmetry 
changing, in agreement with a previously proposed ``generic phase diagram", 
as $d\rightarrow f\rightarrow s$ when $V$ (hence the 
charge fluctuations) are increased. All these are consequences of the structure 
in the charge and spin susceptibilities, which have 
peaks habitating at {\it common or segregated} positions in $k$ space.
\end{abstract}

\pacs{PACS numbers: 74.20.Mn}

\begin{multicols}{2}
\section{INTRODUCTION}
It is gradually becoming clearer that, while spin fluctuations 
are usually considered for electron mechanisms of high-Tc superconductivity, 
charge fluctuations may possibly play essential roles 
in some of the unconventional superconductors.  
Among these are organic metals such as 
$\theta$-(BEDT-TTF)$_2$X\cite{theta} or 
(BEDT-TTF)$_3$Cl$_2$2H$_2$O\cite{lubczynski} that exhibit 
superconductivity sitting adjacent to the charge density wave(CDW) 
in the phase diagram.   This suggests that charge fluctuations can mediate 
pairing as well, just as the spin-fluctuation-mediated pairing\cite{Moriya} 
can appear adjacent to the spin density wave(SDW) 
phase as in the cuprates and 
some of the organic superconductors like $\kappa$-(BEDT-TTF)$_2$X.

Studies for the pairing mediated by charge 
fluctuations have been rather scant.  
Scalapino {\it et al.} have shown, with the random phase 
approximation, that $d_{x^2-y^2}$ pairing gives 
way to $d_{xy}$ in a three-dimensional cubic lattice when charge fluctuations 
become large with the introduction of the off-site 
interaction.\cite{Scalapino3D}
However, systematic studies are yet to come for the charge fluctuations, 
in contrast to the spin-fluctuation mediated 
superconductivity for which favorable situations for its occurrence 
has been extensively discussed\cite{Monthoux,arita}.  
 
There is in fact one proposal in the context of the 
spin-triplet superconductivity 
in a quasi-one-dimensional (1D) organic metal 
(TMTSF)$_2$PF$_6$: Three of the present authors have 
proposed a ``generic phase diagram" in which the dominant pairing symmetry 
changes as $d\rightarrow f\rightarrow s$ as the charge fluctuation 
becomes stronger\cite{Kuroki_cdw}. 
The physical background is as follows: 
triplet pairing is 
very hard to be realized to start with, 
for the reasons identified in Ref.\cite{arita}, 
where the primary one is the strength of the pairing interaction 
for triplets being only 1/3 of that for singlets.  
The situation can be reverted when 
charge fluctuations are dominant as discussed 
in Kuroki {\it et al.}\cite{Kuroki_cdw}, 
but the charge fluctuation was treated 
phenomenologically there, so a microscopic study is highly desirable.  
Namely, while spin fluctuations 
dominate over charge fluctuations when the electron-electron 
interaction is short-ranged (as in the Hubbard model), charge fluctuations 
should become more intense as we increase the 
range of the interaction.  The problem becomes especially intriguing when 
charge and spin fluctuations coexist, since they may 
induce quantum phase transitions among different pairing symmetries.

This is exactly our motivation here to study 
the effect of strong charge fluctuations by adopting 
the extended Hubbard model, as a simplest one in which we can control 
the relative magnitude of the charge fluctuation by varying the off-site 
Coulomb repulsion $V$.  The extended Hubbard
model has been studied, primarily for specific charge densities $n$, 
e.g., half-filling or quarter filling, 
by means of quantum Monte Carlo method\cite{Zhang}, weak coupling 
theory\cite{Tesanovic}, mean-field approximation\cite{Murakami,Seo}, 
second-order 
perturbation\cite{Onozawa}, random phase approximation\cite{Scalapino3D},
fluctuation exchange (FLEX) approximation\cite{esirgen3}, 
slave-boson technique\cite{Merino}, bosonization and 
renormalization group\cite{Sano,Kuroki-Kusakabe}. 
However, the phase diagram of the {\it two-dimensional extended Hubbard model
against $n$ and $V$} has yet to be obtained.

Here we have determined the symmetry of
the dominant pairing in the $V$-$n$ 
space for the extended Hubbard model by focusing on 
the case of isotropic or anisotropic square lattice, 
since many of unconventional superconductors are 
two-dimensional(2D) or quasi-one-dimensional(1D).  
We adopt the FLEX developed by Bickers
{\it et al.}\cite{bickers1,bickers2,dahm,bennemann},
which is a renormalized perturbation scheme to study pairing instabilities when 
exchange of spin and charge fluctuations are considered as dominant diagrams. 
Although this is an approximation, we can explore the 
tendencies for pairing when the system parameters are varied.

We find that (i) there exists a phase transition between 
$d_{xy}$ and $d_{x^2-y^2}$ pairing symmetries, 
(ii) triplet $f$-pairing appears in between the $d_{x^2-y^2}$ and 
CDW phases for intermediate $0.5<n<1.0$ and large $V$, and 
(iii) for anisotropic cases the pairing symmetry 
changes, in agreement with the generic phase diagram\cite{Kuroki_cdw}, 
as $d\rightarrow f\rightarrow s$ when $V$ (hence the 
charge fluctuations) are increased.  
Origin of all these has been identified 
as the structure in the charge and spin susceptibilities, which can have 
peaks habitating at segregated positions in $k$ space.

\section{FORMULATION}
\begin{figure}[h]
\begin{center}
\leavevmode\epsfysize=30mm 
\epsfbox{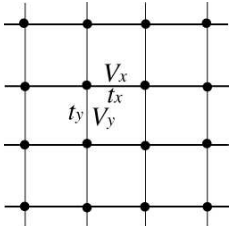}
\caption{A tetragonal lattice with hopping integral $t_x$ along $x$-axis 
and $t_y$ along $y$, along with the nearest-neighbor Coulomb 
repulsion $V_x$ along $x$-axis and  $V_y$ along $y$.}
\label{lattice}
\end{center}
\end{figure}

Let us start with the extended Hubbard Hamiltonian,
\begin{eqnarray}
{\cal H} =&-&\sum_{i,j}^{\rm nn}\sum_{\sigma}t_{ij}c_{i\sigma}^{\dagger}c_{j\sigma}+U\sum_{i}n_{i\uparrow}n_{i\downarrow}\nonumber\\
&+&\frac{1}{2}\sum_{i,j}^{\rm nn}\sum_{\sigma\sigma'}V_{ij}n_{i\sigma}n_{j\sigma'},
\end{eqnarray}

in the standard notation on a tetragonal lattice depicted in Fig. \ref{lattice}.
For the (isotropic) square lattice 
the unit of energy is taken to be the nearest-neighbor $t_{ij}=1.0$, and 
lattice constant $a=1$.

To determine the dominant gap function, we solve 
{\'E}liashberg's equation with the FLEX approximation, 
\begin{eqnarray}
\lambda\phi(k)&=&
-\frac{T}{N}\sum_{k'}\Gamma(k,k')G(k')G(-k')\phi(k')
\label{elia},
\end{eqnarray}
where $\phi$ is the gap function, $G$ Green's function, and 
$\Gamma$ the pairing interaction with $k\equiv (\Vec{k},\omega_n)$. 
The eigenvalue $\lambda$, a measure of the pairing, becomes unity 
at $T=T_C$.  For the calculation 
we take an $N=32\times 32$ lattice, the temperature 
$T=0.02$, and the Matsubara frequency for fermions
$-(2N_c-1)\pi T \leq \omega_n \leq (2N_c-1)\pi T$ with $N_c=1024$.

Esirgen {\it et al.}\cite{esirgen3,esirgen1,esirgen2} have extended 
the FLEX method to general lattice Hamiltonians including 
the extended Hubbard model.  Following them we introduce 
the pairing interaction,
\begin{eqnarray}
\Gamma_{\rm s}(k,k')&=&\nonumber\\
\sum_{\Delta\Vec{r},\Delta\Vec{r}'} \left\{\right.
&\frac{3}{2}&\left[V_{\rm m}\chi_{\rm sp}V_{\rm m}\right](k-k';\Delta\Vec{r};\Delta\Vec{r}')e^{i(\Vec{k}\cdot\Delta\Vec{r}+\Vec{k'}\cdot\Delta\Vec{r}')}\nonumber\\
-&\frac{1}{2}&\left[V_{\rm d}\chi_{\rm ch}V_{\rm d}\right](k-k';\Delta\Vec{r};\Delta\Vec{r}')e^{i(\Vec{k}\cdot\Delta\Vec{r}+\Vec{k'}\cdot\Delta\Vec{r}')} \nonumber\\
+&\frac{1}{2}&V_{\rm s}(0;\Delta\Vec{r};\Delta\Vec{r}')e^{i(\Vec{k}\cdot\Delta\Vec{r}'-\Vec{k'}\cdot\Delta\Vec{r})}
\left.\right\},
\label{s}
\end{eqnarray}
for singlet pairing, and
\begin{eqnarray}
\Gamma_{\rm t}(k,k')&=&\nonumber\\
\sum_{\Delta\Vec{r},\Delta\Vec{r}'} \left\{\right.
-&\frac{1}{2}&\left[V_{\rm m}\chi_{\rm sp}V_{\rm m}\right](k-k';\Delta\Vec{r};\Delta\Vec{r}')e^{i(\Vec{k}\cdot\Delta\Vec{r}+\Vec{k'}\cdot\Delta\Vec{r}')}\nonumber\\
-&\frac{1}{2}&\left[V_{\rm d}\chi_{\rm ch}V_{\rm d}\right](k-k';\Delta\Vec{r};\Delta\Vec{r}')e^{i(\Vec{k}\cdot\Delta\Vec{r}+\Vec{k'}\cdot\Delta\Vec{r}')}\nonumber\\
+&\frac{1}{2}&V_{\rm t}(0;\Delta\Vec{r};\Delta\Vec{r}')e^{i(\Vec{k}\cdot\Delta\Vec{r}'-\Vec{k'}\cdot\Delta\Vec{r})} 
\left.\right\}
\label{t}
\end{eqnarray}
for triplet pairing.  Here $\Delta\Vec{r}(=\Vec{0},\pm\hat{\Vec{x}},\pm\hat{\Vec{y}})$ is 
null or nearest-neighbor vectors, 
\begin{eqnarray}
\chi_{\rm sp} &=& \overline{\chi}/(1+V_{\rm m}\overline{\chi}), \nonumber\\
\chi_{\rm ch} &=& \overline{\chi}/(1+V_{\rm d}\overline{\chi}) \nonumber
\end{eqnarray}
are the spin and charge susceptibilities, respectively, where 
$\overline{\chi}$ is the irreducible susceptibility,
\begin{eqnarray}
\overline{\chi}(q;\Delta\Vec{r};\Delta\Vec{r}')=-\frac{T}{N}\sum_{k'}e^{i\Vec{k'}\cdot(\Delta\Vec{r}-\Delta\Vec{r}')}G(k'+q)G(k'),
\end{eqnarray}
and $V_{\rm d} (V_{\rm m})$ is the coupling between density (magnetic) 
fluctuations,
{\small
\begin{eqnarray}
&V_{\rm d}&(q;\Delta\Vec{r};\Delta\Vec{r}')\nonumber\\
&=& 
\left\{
\begin{array}{ll}
U+4[V_x\cos(q_x)+V_y\cos(q_y)], & \Delta\Vec{r}=\Delta\Vec{r}'=\Vec{0},\\
-V_x, & \Delta\Vec{r}=\Delta\Vec{r}'=\pm\hat{\Vec{x}}\\
-V_y, & \Delta\Vec{r}=\Delta\Vec{r}'=\pm\hat{\Vec{y}}\\
\end{array}
\right.\\
&V_{\rm m}&(q;\Delta\Vec{r};\Delta\Vec{r}')=
\left\{
\begin{array}{ll}
-U, & \Delta\Vec{r}=\Delta\Vec{r}'=\Vec{0},\\
-V_x, & \Delta\Vec{r}=\Delta\Vec{r}'=\pm\hat{\Vec{x}}\\
-V_y, & \Delta\Vec{r}=\Delta\Vec{r}'=\pm\hat{\Vec{y}},\\
\end{array}
\right.
\end{eqnarray}
}
where $q\equiv(\Vec{q},\epsilon_n)$ with $\epsilon_n=2n\pi T$ 
being the Matsubara frequencies for bosons.
We have found that the $q$ dependence 
of $V_{\rm m}$ and $V_{\rm d}$ does not in fact affect $\Gamma_{\rm s}$ and
$\Gamma_{\rm t}$ significantly.  
Accordingly the peak position of $\chi_{\rm ch}$ is almost the same as that for 
$V_{\rm d}\chi_{\rm ch}V_{\rm d}$ term in the expression for $\Gamma$.

$V_{\rm s}(0;\Delta\Vec{r};\Delta\Vec{r}')$, $V_{\rm t}(0;\Delta\Vec{r};\Delta\Vec{r}')$, appearing in the last lines in eqs.(\ref{s},\ref{t}) respectively, 
are constant terms involving $U$ and $V$,
\begin{eqnarray}
V_{\rm s}(\Vec{q};\Delta\Vec{r};\Delta\Vec{r}')&=&
\left\{
\begin{array}{rl}
2U, & \Delta\Vec{r}=\Delta\Vec{r}'=\Vec{0},\\
V_x, & \Delta\Vec{r}=\Delta\Vec{r}'=\pm\hat{\Vec{x}}\\
V_xe^{\pm iq_x}, &\Delta\Vec{r}=-\Delta\Vec{r}'=\pm\hat{\Vec{x}}\\
V_y, & \Delta\Vec{r}=\Delta\Vec{r}'=\pm\hat{\Vec{y}}\\
V_ye^{\pm iq_y}, &\Delta\Vec{r}=-\Delta\Vec{r}'=\pm\hat{\Vec{y}}\\
\end{array}
\right.\\
V_{\rm t}(\Vec{q};\Delta\Vec{r};\Delta\Vec{r}')&=&
\left\{
\begin{array}{rl}
V_x, & \Delta\Vec{r}=\Delta\Vec{r}'=\pm\hat{\Vec{x}}\\
-V_xe^{\pm iq_x}, &\Delta\Vec{r}=-\Delta\Vec{r}'=\pm\hat{\Vec{x}}\\
V_y, & \Delta\Vec{r}=\Delta\Vec{r}'=\pm\hat{\Vec{y}}\\
-V_ye^{\pm iq_y}, &\Delta\Vec{r}=-\Delta\Vec{r}'=\pm\hat{\Vec{y}}\\
\end{array}
\right.
\end{eqnarray}
 
When the off-site interaction $V$ is introduced 
all the vertices ($V_{\rm m}$, $V_{\rm d}$, $V_{\rm s}$, $V_{\rm t}$) 
as well as the susceptibilities become 
$(Z+1)\times (Z+1)$ matrices for the lattice coordination number $Z 
(=4$ for the square lattice).

\section{RESULT}

\subsection{Square lattice}
\begin{figure}[h]
\begin{center}
\leavevmode\epsfysize=50mm 
\epsfbox{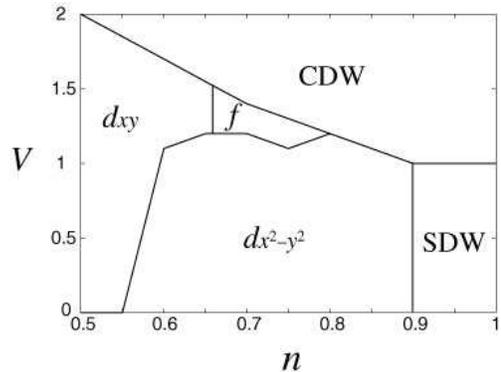}
\caption{Phase diagram against $V$ and $n$ with 
$U/t=4$ for the 2D extended Hubbard model.  
}
\label{phase}
\end{center}
\end{figure}

Let us first display the obtained phase diagram against $V$ and 
the band filling $n$ for the square lattice in Fig.\ref{phase}.  
The phase diagram is drawn by assuming 
that the pairing instability that has the largest
$\lambda$ in {\'E}liashberg's equation, calculated at $T=0.02$ here, 
has the highest transition temperature, 
because, with the present complexity of the model, 
it is difficult to extend the FLEX calculation to lower temperatures.  
While the value 
of $\lambda$ for $T=0.02$ is still much smaller than unity (see below), 
it is expected that the order in which the values of $\lambda$ for 
various phases appear do not change for $T \rightarrow 0$.   
The density waves (CDW or SDW) are identified as the 
region where the respective (charge or spin) 
susceptibility diverges (at $T>0.02$ in the present calculation).  
The area of the CDW and SDW phases 
may expand at lower temperatures, but the phase diagram is
not expected to change significantly. 

Figure \ref{phase} is thus obtained, with the on-site Coulomb repulsion 
fixed at $U=4$ hereafter, and we immediately note 
that CDW, SDW, singlet superconductivity(SC), and triplet SC 
all appear on the $(V,n)$ plane.  
The phase diagram is reminiscent of 
that for one-dimensional extended Hubbard model obtained 
with the Tomonaga-Luttinger theory\cite{TL}.  
However, an essential difference is that 
$U$ or $V$ has to be negative (attractive) to realize 
SC phases in one dimension, while we are talking about 
the case when both $U$ and $V$ are repulsive in 2D.   
Another comment is that in our calculation we cannot treat the Mott
  insulator, which should appear at n close enough to the half-filling $(n=1)$.

Before elaborating on the superconducting phases, let us make a remark on 
density waves.  
Intuitively, CDW should appear for strong enough $V$, while 
we should have SDW for the band filling 
close enough to the half-filling.  
The boundary between CDW and SDW for $n\rightarrow 1$ is 
seen in the present result to fall upon a line 
representing $V=1$, 
which agrees with a mean-field picture: 
In the CDW state where electrons doubly-occupy every 
other sites, each electron feels on average 
an on-site energy $U/2$ per electron, 
while in the SDW state each electron, singly-occupied, feels an off-site 
repulsion $V/2 \times Z =2V$, so the 
SDW/CDW boundary corresponds to $V=U/4$.\cite{Zhang} 
As for the wave vectors describing the CDW (SDW) ordering, 
they are as indicated by the peaks (around $(\pi,\pi)$) 
in the charge (spin) susceptibilities, which are 
displayed below.

\begin{figure}[h]
\begin{center}
\leavevmode\epsfysize=50mm 
\epsfbox{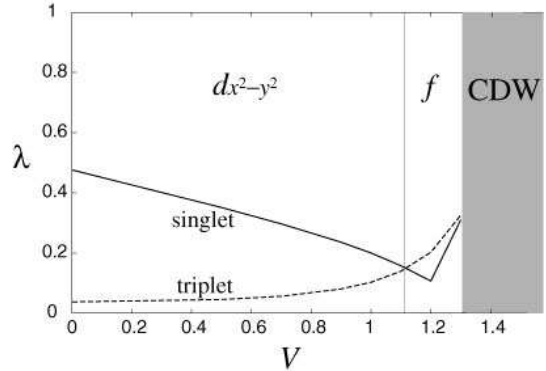}
\caption{
The maximum eigenvalue, $\lambda$, of {\'E}liashberg's equation 
for the spin-singlet (solid line) and triplet (dotted) channel 
as a function of $V$ for $n=0.7$ with the dominant orbital symmetry indicated. 
The CDW phase is identified from the divergence in the charge 
susceptibility.
}
\label{fst}
\end{center}
\end{figure}

\begin{figure}[h]
\begin{center}
\leavevmode\epsfysize=40mm 
\epsfbox{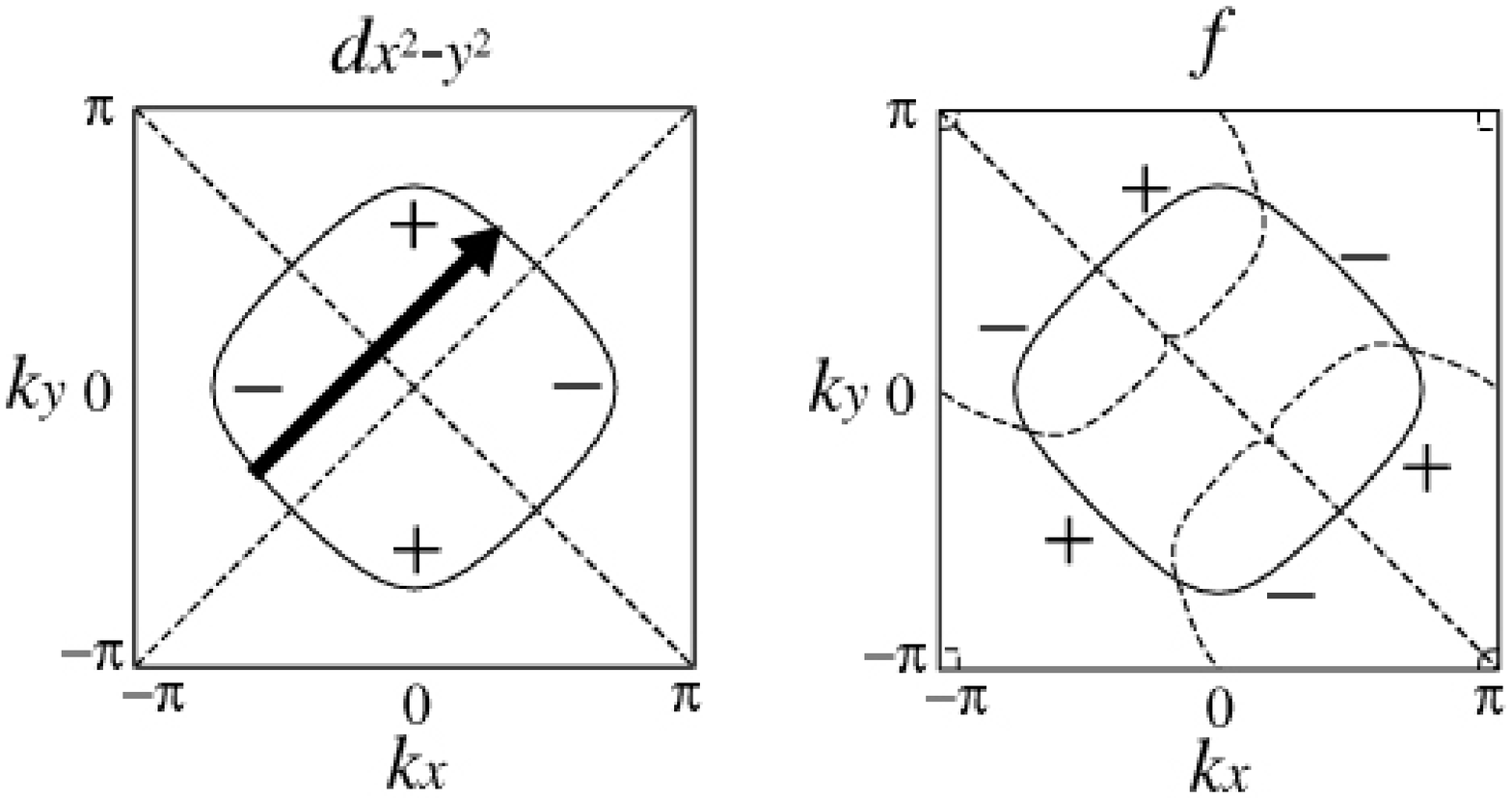}
\caption{
The $f$-wave (irreducible representation $\Gamma^{-}_{5}$) gap function
in $k$ space (right panel) for $n=0.7$ and $V=1.3$, 
as compared with the $d_{x^2-y^2}$ (left) 
for $n=0.7$, $V=0.5$. 
Nodal lines are represented by dotted lines, while 
the Fermi surface by solid curve on which the 
sign of the gap function is indicated. 
The arrow indicates a typical scattering process mediated 
by spin fluctuations.
}
\label{f-wave}
\end{center}
\end{figure}

A salient feature for superconducting phases in 
Fig. \ref{phase} is that a spin-triplet $f$-phase 
appears just below the CDW phase 
for an intermediate region of $n$ and $V$.  The behavior of $\lambda$ 
when $V$ is varied with a fixed $n=0.7$ 
is depicted in Fig.\ref{fst}.\cite{kink} 
Figure \ref{f-wave} shows the gap function in $k$ space 
for the $f$-wave, which should be called, more precisely, 
$\Gamma^{-}_{5}$ in the group theoretical representation. 
This has only one nodal line 
passing $\Gamma$ point, but may be called $f$ in that 
the gap function ($\sim {\rm sin}(k_x)+{\rm sin}(k_y)+
{\rm const.}[{\rm sin}(2k_x)+{\rm sin}(2k_y)]$) 
changes sign as $+-+-+-$ around the Fermi surface.\cite{commentf}

\subsection{Physical origin --- common vs segregated peaks in $\chi_{\rm sp}$ 
and $\chi_{\rm ch}$}

\begin{figure}[h]
\begin{center}
\leavevmode\epsfysize=70mm 
\epsfbox{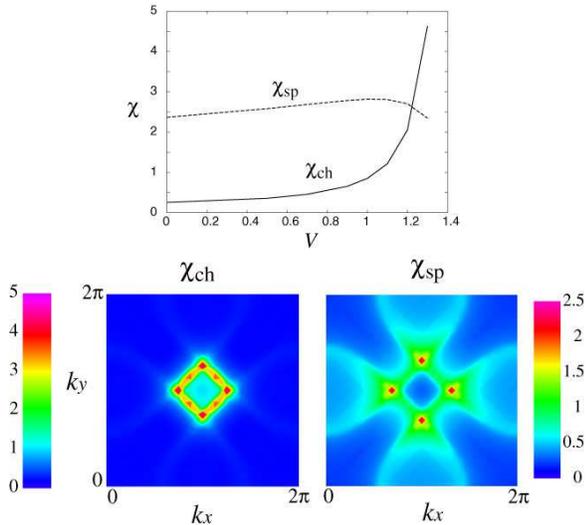}
\caption{
Top: the maximum eigenvalue of the charge (solid curve) and spin 
(dotted) susceptibilities as a function of $V$ for $n=0.7$. 
Bottom: color-coded plots of the 
charge (left) and spin (right) susceptibilities in the $(k_x,k_y)$ space 
for $V=1.3$ and $n=0.7$.   
The susceptibilities [$\sim$ (energy)$^{-1}$] in units 
in which $t=1$.  
}
\label{csdwf}
\end{center}
\end{figure}

We can keep track of the origin of the spin-triplet instability 
by looking at the structure 
(peak intensities and peak positions in $k$ space) of the charge 
($\chi_{\rm ch}$) and spin ($\chi_{\rm sp}$) susceptibilities 
(static with Matsubara frequency$=0$) 
in Fig. \ref{csdwf}.  
The peak intensity of $\chi_{\rm ch}$ is seen to exceed 
that of $\chi_{\rm sp}$ as $V$ is increased.   
If we turn to the structure in $k$ space, $\chi_{\rm ch}$ and 
$\chi_{\rm sp}$ have similar peak positions. 
If we go back to eqs.(\ref{s}),(\ref{t}), the 
spin-fluctuation mediated pairing interaction (the first 
line on the right-hand side) and the charge-fluctuation 
mediated one (the second) act destructively 
(with opposite signs for the two terms in eq.(\ref{s})) 
for singlet pairing, whereas 
they act {\it constructively} for triplet pairing (same signs 
in eq.(\ref{t})).  So we can expect the realization of triplet pairing 
($|\Gamma_{\rm t}|>|\Gamma_{\rm s}|$) 
for large enough charge-fluctuation 
mediated pairing interactions (i.e., for large enough $V$) 
{\it provided $\chi_{\rm ch}$ and 
$\chi_{\rm sp}$ have common peak positions}.

Physically, $\chi_{\rm sp}=\overline{\chi}/(1-U\overline{\chi})$ 
while $\chi_{\rm ch}=\overline{\chi}/(1+U\overline{\chi})$ 
when $V=0$, so the nesting of the Fermi surface (which 
dominates $\overline{\chi}$) determines the peaks of $\chi_{\rm sp}$ 
(but {\it not} those of $\chi_{\rm ch}$).  When $V$ is switched on, 
$\chi_{\rm sp}=\overline{\chi}/(1+V_{\rm m}\overline{\chi})$ 
and $\chi_{\rm ch}=\overline{\chi}/(1+V_{\rm d}\overline{\chi})$ 
may have common peak positions, which is what is happening here.

\begin{figure}[h]
\begin{center}
\leavevmode\epsfysize=40mm 
\epsfbox{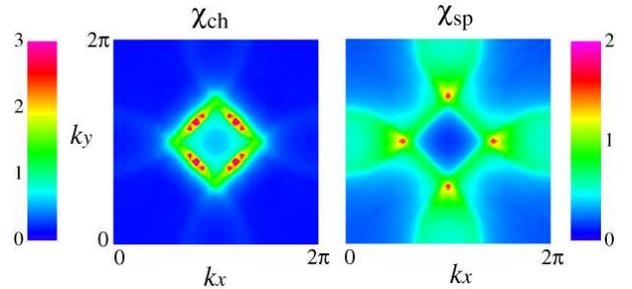}
\caption{
A plot similar to Fig. \ref{csdwf}, for $n=0.6$ with $V=1.6$.
}
\label{csdw}
\end{center}
\end{figure}

How the situation is altered for smaller $n$ 
for which $d_{xy}$ pairing turns out to be dominant?  In Fig.\ref{csdw}, 
which shows $\chi_{\rm ch}$ and $\chi_{\rm sp}$ for $n=0.6$ with $V=1.6$, 
$\chi_{\rm ch}$ has a larger peak than 
$\chi_{\rm sp}$ does, but the peak positions of 
$\chi_{\rm ch}$ are {\it distinct} from those of $\chi_{\rm sp}$. 
Namely, while $\chi_{\rm ch}$ has peaks at four points around 
$\Vec{Q}_{\rm ch} \simeq (\pm \pi, \pm \pi)$, $\chi_{\rm sp}$ has 
peaks that are shifted toward 
$\Vec{Q}_{\rm sp}=(\pm \pi,0)$ and $(0, \pm \pi)$. 
Now, as a basic property of {\'E}liashberg's 
eq.(\ref{elia}) with eq.(\ref{s}) plugged, 
to realize a large $\lambda$ requires that 
(i) the gap function $\phi$ must change sign 
(across the typical pair-scattering momentum transfer $\Vec{Q}_{\rm sp}$) 
to turn the originally repulsive spin-fluctuation-mediated 
interaction (i.e., the 
$V_{\rm m}\chi_{\rm sp}V_{\rm m}$ term with a positive coefficient, $3/2$ 
in eq.(\ref{s})) into an effective attraction in the gap equation, while 
(ii) $\phi$ must not change sign 
(across the pair-scattering momentum transfer $\Vec{Q}_{\rm ch}$) 
to make the originally attractive charge-fluctuation-mediated 
interaction (the 
$V_{\rm d}\chi_{\rm ch}V_{\rm d}$ term with a negative coefficient, $-1/2$) 
remain attractive.

\begin{figure}[h]
\begin{center}
\leavevmode\epsfysize=40mm 
\epsfbox{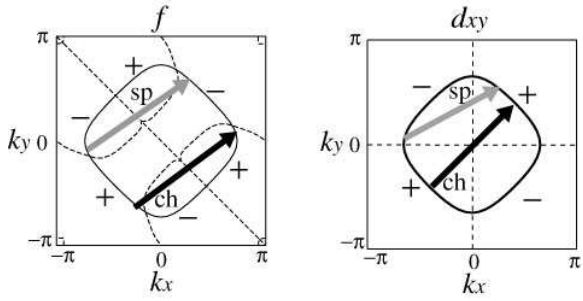}
\caption{
The Fermi surface for $n=0.6$, $V=1.6$ 
with $d_{xy}$ pairing(right), and for $n=0.7$, $V=1.3$  
with the $f$ pairing(left). 
The black (gray) arrows indicate typical scattering processes mediated 
by charge (spin) fluctuations.
}
\label{peak}
\end{center}
\end{figure}

We can see in Fig.\ref{peak} that $d_{xy}$ pairing does satisfy 
the above condition. 
For $n$ closer to half-filling, by contrast, $\chi_{\rm sp}$ becomes 
dominant with $\Vec{Q}_{\rm sp}\simeq (\pm \pi, \pm \pi)$ 
(an arrow in Fig.\ref{f-wave}), 
so $d_{x^2-y^2}$ is favored as usual.  
So the picture obtained here is that the pairing symmetry can 
change, even within the spin-singlet channel, 
when common peaks between $\chi_{\rm sp}$ and 
$\chi_{\rm ch}$ change into 
{\it segregated} peaks (as $n$ and/or $V$ are changed).  
This is contrasted with the above case of 
triplet $f$, for which a transition between 
singlet and triplet ($d \leftrightarrow f$) can occur with 
the common peaks between $\chi_{\rm sp}$ and 
$\chi_{\rm ch}$ throughout.   
For $n=0.5$, Merino {\it et al.}\cite{Merino} 
and Kobayashi {\it et al.}\cite{Kobayashi_Tanaka} have recently derived
similar results using the slave-boson technique and RPA, respectively
which are consistent with the present result.

\subsection{Quasi one-dimensional lattice}

Let us finally discuss the anisotropic (quasi 1D) case.
While the original proposal made in 
Ref.\cite{Kuroki_cdw} for the triplet `$f$-wave' pairing 
in (TMTSF)$_2$X (where the symmetry refers to the warped Fermi surface; 
see Fig.\ref{1Dsu} below) 
is for a quarter-filled case, 
the competition between the charge and spin fluctuations should become 
more stringent near half-filling, where $U$ introduces 
$2k_F$ spin fluctuations while $V$ enhances $2k_F$ 
charge fluctuations. Thus the 
triplet pairing should dominate over singlet $d$ beyond some value of $V$.  
Note that the triplet gap function has to be $f$ rather 
than $p$, since, with $\Gamma_{\rm t}(\Vec{Q})$ negative (attractive), 
the gap has to have the same sign across the nesting vector $\Vec{Q}$.
When $V$ becomes 
even larger, $\Gamma_{\rm s}(\Vec{Q})$ turns negative (attractive), so that 
the singlet $s$ with no nodes on the Fermi surface 
will take over.

In order to show that this prediction is indeed realized, we have performed 
a FLEX calculation on a quasi 1D lattice with $t_x=1.0, t_y=0.2$. 
To represent the quasi 1D system we have here 
assumed that the off-site repulsion only acts between 
nearest neighbors along $x$ (the conductive direction), 
although this assumption does not qualitatively alter the result.
Figure \ref{1Dsu} plots 
$\lambda$ as a function of $V_x$ for $n=0.9$, along with 
the forms of $d$, $f$, and $s$ gap functions.   
We can see that the dominant pairing 
changes as $d\rightarrow f\rightarrow s$ with $V$, 
in an exact agreement with Ref.\cite{Kuroki_cdw}.

\begin{figure}[h]
\begin{center}
\leavevmode\epsfysize=80mm 
\epsfbox{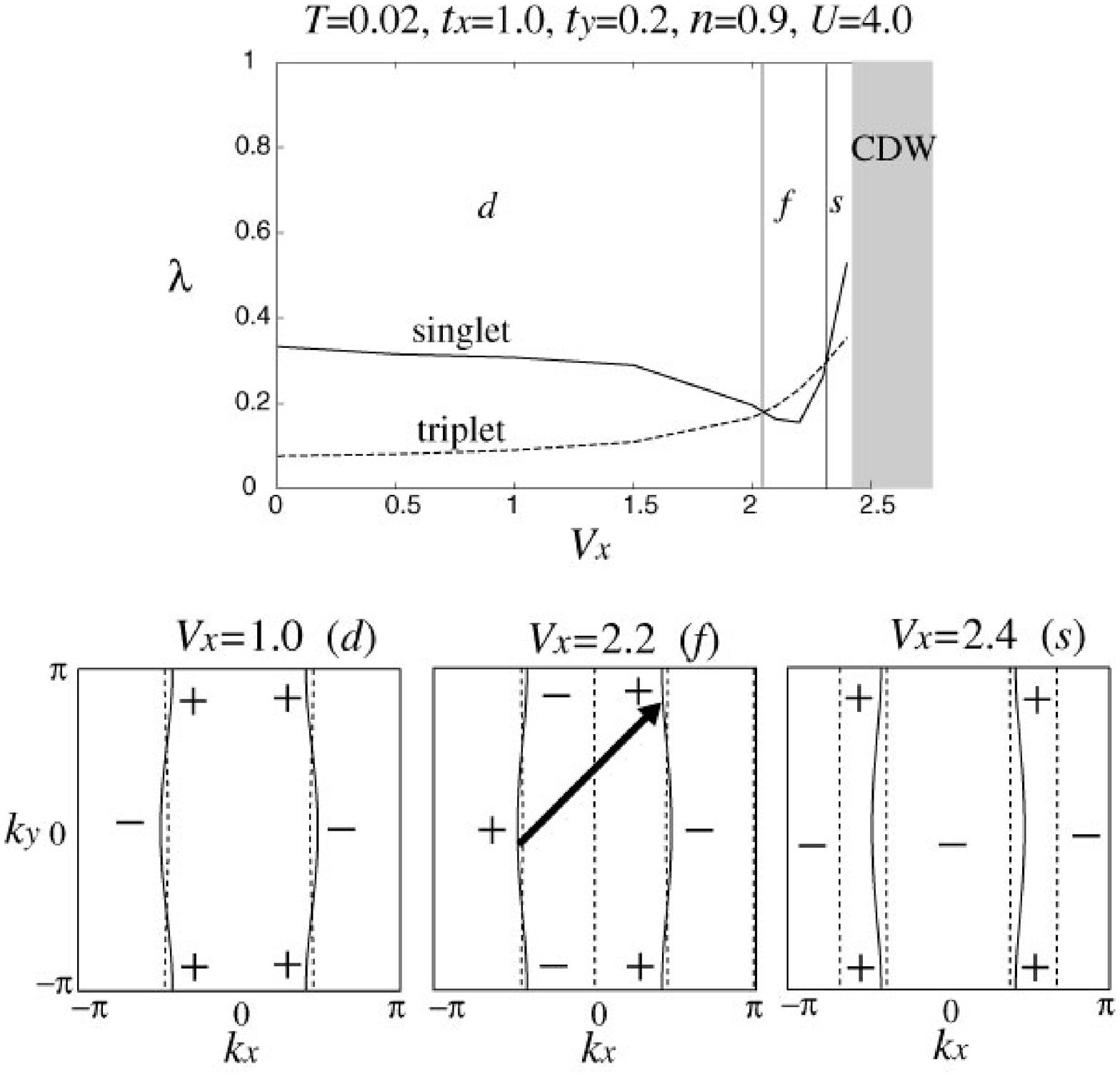}
\caption{Result for a quasi 1D system with $t_x=1.0, t_y=0.2$ for $n=0.9$.  
Top: a plot similar to Fig.\ref{fst}. 
Bottom: The dominant gap function (with nodal lines represented 
by dotted lines) for $V_x=1.0$ ($d$ wave), $V_x=2.2$ ($f$), 
and $V_x=2.4$ ($s$), along with the Fermi surface
(solid curves) on which an arrow represents the nesting vector.
}
\label{1Dsu}
\end{center}
\end{figure}

\section{CONCLUSION}
We have obtained the phase diagram for the 2D extended 
Hubbard model with the FLEX approximation. We have found 
that $f$-wave pairing is favored for intermediate $n$ and large $V$.
In the more dilute regime ($n\sim 0.5$) 
the peak position of $\chi_{\rm ch}$ and 
$\chi_{\rm sp}$ is separated, and $d_{x^2-y^2}$ pairing gives 
way to $d_{xy}$.  These are the consequences of how each 
pairing symmetry can exploit charge and spin 
fluctuations, and we end up with a 
picture that the pairing symmetry can 
change, even within the spin-singlet channel ($d_{x^2-y^2} 
\leftrightarrow d_{xy}$), 
when common peaks between $\chi_{\rm sp}$ and 
$\chi_{\rm ch}$ change into 
segregated peaks (as $n$ and/or $V$ are changed), which 
is contrasted with the case of common peaks throughout 
with a transition triplet $f \leftrightarrow d$.   
In the anisotropic (quasi-one-dimensional) case, the dominant 
pairing gap function changes as $d\rightarrow f\rightarrow s$ 
with $V$, which agrees with the phenomenological theory\cite{Kuroki_cdw}.

In a broad context it should be interesting to examine how 
   $V$ (hence the charge fluctuation 
   mediated interaction) can dominate the pairing symmetry in 
   unconventional superconductors, although the nature of 
   the pairing may be sensitively affected by the underlying 
   band structure of each material.

\section{ACKNOWLEDGMENTS}
We would like to thank Masao Ogata for illuminating discussions.
Numerical calculations were performed at the supercomputer center, ISSP.

\end{multicols}

\begin{references}
\bibitem{commentf} The true ground state may be $f+if$, since 
Fig.\ref{f-wave} rotated by 90 degrees exists as a degenerate state.
\end{references}
\end{document}